\documentstyle[12pt]{article}

\def\gsim{\ \rlap{\raise 3pt \hbox{$>$}}{\lower 3pt \hbox{$\sim$}}\ }
\def\lsim{\ \rlap{\raise 3pt \hbox{$<$}}{\lower 3pt \hbox{$\sim$}}\ }

\newcommand\ijmpa[3]{Int.\ J.\ Mod.\ Phys.\ {\bf A#1} (#2) #3}

\newcommand\plb[3]{Phys.\ Lett.\ {\bf B#1} (#2) #3} 
\newcommand\prd[3]{Phys.\ Rev.\ {\bf D#1} (#2) #3}
\newcommand\prl[3]{Phys.\ Rev.\ Lett.\ {\bf #1} (#2) #3}
\newcommand\rmp[3]{Rev.\ Mod.\ Phys.\ {\bf #1} (#2) #3}
\newcommand\zpc[3]{Z.\ Phys.\ {\bf C#1} (#2) #3}
\newcommand{\hepph}[1]{{\tt hep-ph/#1}}

\begin{document}

\begin{titlepage}

\begin{flushright}
CERN-TH/98-273\\ 
EFI-98-38\\
hep-ph/9808493
\end{flushright}

\vspace{1.0cm}
\begin{center}
\Large\bf \boldmath  New Bound on $\gamma$ from $B^\pm\to\pi K$ Decays
\unboldmath
\end{center}

\vspace{0.5cm}
\begin{center}
Matthias Neubert\\[0.1cm] 
{\sl Theory Division, CERN, CH-1211 Geneva 23, Switzerland} \\[0.3cm] 
and\\[0.3cm] 
Jonathan L. Rosner\\[0.1cm]
{\sl Enrico Fermi Institute and Department of Physics\\ 
University of Chicago, Chicago, IL 60637, USA}
\end{center}

\vspace{0.5cm}
\begin{abstract}
\vspace{0.2cm}\noindent  A bound on the angle $\gamma$ of the
unitarity triangle is derived using experimental information on the
CP-averaged branching ratios for the rare decays $B^\pm\to\pi^\pm K^0$
and $B^\pm\to\pi^0 K^\pm$. The theoretical description is cleaner than
the  Fleischer--Mannel analysis of the decays $B^\pm\to\pi^\pm K^0$
and $B^0\to\pi^\mp K^\pm$ in that the two decay rates differ only in a
single isospin amplitude, which has a simple structure in the SU(3)
limit. As a consequence, electroweak penguin contributions and strong
rescattering  effects can be taken into account in a model-independent
way. The  resulting bound excludes values of $\cos\gamma$ around
$\approx 0.6$ and is thus largely complementary to indirect
constraints derived from a global analysis of the unitarity triangle.
\end{abstract}

\vspace{0.5cm} 
\centerline{(To appear in Physics Letters B)}

\vfil
\noindent
August 1998

\end{titlepage}

The study of CP violation in the weak decays of $B$ mesons is the main
target of present and future $B$ factories. It will provide stringent
tests of the flavor sector of the Standard Model and  of the CKM
paradigm, according to which all CP violation results  from the
presence of a single complex phase in the quark mixing matrix. The
precise determination of the sides and angles of the  unitarity
triangle, which is a graphical representation of the  unitarity
relation $V_{ub}^* V_{ud}+V_{cb}^* V_{cd}+V_{tb}^* V_{td}=0$, plays 
a central role in this program \cite{review}. The angle 
$\beta=-\mbox{Arg}(V_{tb}^* V_{td})$ will be accessible at the 
first-generation $B$ factories through the measurement of CP
violation in the decay $B\to J/\psi K_S$. The angle 
$\gamma=\mbox{Arg}(V_{ub}^* V_{ud})$ is, however, much harder to 
determine. The sum $(\beta+\gamma)$ can, in principle, be extracted in 
a theoretically clean way from measurements of CP violation in the 
decays $B\to\pi\pi$ (or in the related decays $B\to\pi\rho$ and 
$\rho\rho$), but because of experimental difficulties such as the
detection of the mode $B\to\pi^0\pi^0$ this will be a long-term 
objective.

Fleischer and Mannel have argued that some information on the angle
$\gamma$ can be derived from the measurement of the  branching ratios
for the decays $B^\pm\to\pi^\pm K^0$ and  $B^0\to\pi^\mp K^\pm$,
averaged over CP-conjugate modes \cite{FM}. In their original work,
they obtained the bound
\begin{equation}
   R = \frac{\tau(B^+)}{\tau(B^0)}\,  \frac{\mbox{Br}(B^0\to\pi^- K^+)
       + \mbox{Br}(\bar B^0\to\pi^+ K^-)} {\mbox{Br}(B^+\to\pi^+ K^0)
       + \mbox{Br}(B^-\to\pi^-\bar K^0)} \ge \sin^2\!\gamma \,,
\label{FMbound}
\end{equation}
which would exclude a region around $\gamma=90^\circ$ provided that
$R<1$, a possibility allowed by the first measurement of this ratio
yielding $R=0.65\pm 0.40$ \cite{CLEOR}. It has later been realized,
however, that the bound (\ref{FMbound}) may be subject to sizable
corrections arising from final-state interactions and electroweak
penguin contributions, which are difficult to quantify in a
model-independent way \cite{GRKpi}--\cite{At97}. To control such
effects would  require a more sophisticated analysis using
information from other decays such as $B\to K\bar K$
\cite{Robert,GRKK}.  In addition to these theoretical obstacles, the
prospects for deriving useful information on $\gamma$ using the ratio
$R$ are reduced by the fact that the CLEO Collaboration has announced
a new measurement of this quantity yielding $R=1.0\pm 0.4$
\cite{CLEOnew}.\footnote{Some information on $\gamma$ can be obtained 
even when $R\approx 1$ if one uses separate measurements of the decay 
rates for $B^0\to\pi^- K^+$ and $\bar B^0\to\pi^+ K^-$ 
\protect\cite{GRKpi}.}

In the present note we propose a variant of the Fleischer--Mannel
analysis, which is theoretically cleaner and implies a non-trivial  
constraint on $\gamma$ provided present, preliminary data are 
confirmed by future measurements. Electroweak penguin contributions 
play an important role in this analysis, but they can be
controlled in a model-independent way. Consider the ratio
\begin{equation}
   R_* = \frac{\mbox{Br}(B^+\to\pi^+ K^0) + \mbox{Br}(B^-\to\pi^-\bar
               K^0)} {2[\mbox{Br}(B^+\to\pi^0 K^+) +
               \mbox{Br}(B^-\to\pi^0 K^-)]} \equiv (1-\Delta_*)^2 \,,
\label{Rstar}
\end{equation}
which would approach unity in the limit where  $b\to s\bar q q$
penguin transitions involving top- or charm-quark loops dominate over
the Cabibbo-suppressed $b\to u\bar u s$ transitions.  Deviations from
this limit are measured by the quantity $\Delta_*$. The CLEO
Collaboration has recently reported the preliminary results
$\mbox{Br}(B^\pm\to\pi^\pm K^0)=(1.4\pm 0.5\pm 0.2)\times 10^{-5}$ and
$\mbox{Br}(B^\pm\to\pi^0 K^\pm)=(1.5\pm 0.4\pm 0.3)\times 10^{-5}$ for
the CP-averaged branching ratios \cite{CLEOnew}, which imply
\begin{equation}
   R_*^{\rm exp} = 0.47\pm 0.24 \,, \qquad  \Delta_*^{\rm exp} =
0.32\pm 0.17 \,,
\label{Rstexp}
\end{equation}
corresponding to a deviation of $R_*$ from unity of about two standard
deviations. The goal of this note is to transform this measurement
into a model-independent constraint on the angle $\gamma$.

The effective weak Hamiltonian governing the decays $B\to\pi K$ has
the form \cite{Heff}
\begin{equation}
   {\mathcal H} = \frac{G_F}{\sqrt 2}\,\bigg\{ \sum_{i=1,2} C_i \Big(
   \lambda_u\,Q_i^u + \lambda_c\,Q_i^c \Big) - \lambda_t
   \sum_{i=3}^{10} C_i\,Q_i  \bigg\} + \mbox{h.c.} \,,
\end{equation}
where $\lambda_q=V_{qb}^* V_{qs}$ are products of CKM matrix elements
satisfying the unitarity relation $\lambda_u+\lambda_c+\lambda_t=0$,
$C_i$ are Wilson coefficients, and $Q_i$ are local four-quark
operators.  Relevant to the further discussion are the isospin quantum
numbers of these operators. The current--current operators
$Q_{1,2}^u\sim\bar b s\bar u u$ have components with $\Delta I=0$ and
$\Delta I=1$; the current--current operators  $Q_{1,2}^c\sim\bar b
s\bar c c$ and the QCD penguin operators $Q_{3,\dots,6} \sim \bar b
s\sum\bar q q$ have $\Delta I=0$; the electroweak penguin operators
$Q_{7,\dots,10}\sim\bar b s \sum e_q\bar q q$, where $e_q$ are the
electric charges of the quarks, have $\Delta I=0$ and $\Delta
I=1$. Since the initial $B$ meson has $I=\frac 12$ and the final
states $(\pi K)$ can be decomposed into components with $I=\frac 12$
and $I=\frac 32$, the physical $B\to\pi K$ decay amplitudes can be
described in terms of three isospin amplitudes. They are called
$B_{1/2}$, $A_{1/2}$, and $A_{3/2}$ referring, respectively, to
$\Delta I=0$ with $I_{\pi K}=\frac 12$, $\Delta I=1$ with $I_{\pi
K}=\frac 12$, and  $\Delta I=1$ with $I_{\pi K}=\frac 32$
\cite{Ne97,Gron,NQ}.  From the isospin decomposition of the effective
Hamiltonian it is obvious which operator matrix elements and weak
phases enter the various isospin amplitudes. The resulting expressions
for the decay amplitudes relevant to our discussion are
\begin{eqnarray}
   {\mathcal A}(B^+\to\pi^+ K^0)
   &=& B_{1/2} + A_{1/2} + A_{3/2} \,, \nonumber\\
   - \sqrt 2\,{\mathcal A}(B^+\to\pi^0 K^+)
   &=& B_{1/2} + A_{1/2} - 2 A_{3/2} \,, \nonumber\\
   - {\mathcal A}(B^0\to\pi^- K^+)
   &=& B_{1/2} - A_{1/2} - A_{3/2} \,.
\end{eqnarray}
Experimental data as well as theoretical expectations indicate that
the amplitude $B_{1/2}$, which includes the contributions of the QCD
penguin operators, is an order of magnitude larger than the amplitudes
$A_{1/2}$ and $A_{3/2}$ \cite{Ne97,GrRo}.  Yet, the fact that
$A_{1/2}$ and $A_{3/2}$ are different from zero is responsible for the
deviations of the ratios $R$ and $R_*$ from unity.  An important
advantage of our analysis is that the two decay amplitudes entering
the ratio $R_*$ in (\ref{Rstar}) differ only in the sign of the
isospin amplitude $A_{3/2}$, which in the SU(3) limit is related to a
matrix element of a single combination of local four-quark operators
in the effective weak Hamiltonian.  Despite the presence of
contributions carrying different weak phases,  there is thus a single
strong-interaction phase associated with  this amplitude. This fact
will allow us to control rescattering effects  and calculate
electroweak penguin contributions in a model-independent way.  On the
contrary, the decay amplitudes entering the ratio $R$ considered in
the original Fleischer--Mannel analysis differ in the signs of the two
isospin amplitudes $A_{1/2}$ and $A_{3/2}$, introducing hadronic
uncertainties into the calculation \cite{GRKpi}--\cite{At97}.

Taking advantage of the fact that the top- and charm-quark penguin
contributions to $B_{1/2}$ are much larger than all other
contributions to the isospin amplitudes, we write $B_{1/2}=|P_{ct}| 
e^{i\phi_P} [e^{i\pi} + O(\varepsilon)]$, where $e^{i\phi_P}$ is a
strong-interaction phase, and $e^{i\pi}$ is the weak phase associated
with the top-quark penguin contribution. We generically denote by 
$O(\varepsilon)$ contributions arising from the quark decay 
$\bar b\to\bar u u\bar s$, which are Cabibbo-suppressed relative to 
the leading penguin amplitudes and carry the weak phase $e^{i\gamma}$ 
of the parameter $\lambda_u$. Our main result is that the isospin 
amplitude $A_{3/2}$, which causes the deviation of the ratio $R_*$ 
from unity, can be written as
\begin{equation}
   \frac{3 A_{3/2}}{|P_{ct}|} = - \varepsilon_{3/2}\,e^{i\phi_{3/2}}
   (e^{i\gamma} - \delta_{\rm EW}) \,,
\label{ratio}
\end{equation}
where $e^{i\phi_{3/2}}$ is a strong-interaction phase, and 
$\delta_{\rm EW}$ is to a good approximation a real parameter 
accounting for the contributions of electroweak penguin operators. 
The fact that this parameter does not carry a non-trivial 
strong-interaction  phase is crucial and will be derived below. We 
define the parameter $\varepsilon_{3/2}$ to be real and positive. In 
the diagrammatic amplitude approach of flavor-flow topologies, 
$\varepsilon_{3/2}=|T+C|/|P_{ct}|$ is the ratio of color-allowed
plus color-suppressed tree amplitudes to the sum of the top- and
charm-quark penguin amplitudes \cite{Chau}--\cite{ampl}.  Using the
parametrization (\ref{ratio}), we obtain for the quantity $\Delta_*$
defined in (\ref{Rstar}) the result
\begin{equation}
   \Delta_* = - \varepsilon_{3/2} \cos\Delta\phi\,
   (\cos\gamma - \delta_{\rm EW}) + O(\varepsilon^2) \,,
\label{Delstar}
\end{equation}
where $\Delta\phi=\phi_{3/2}-\phi_P$.  The phase conventions adopted
above take into account the naive phase relation between penguin and
tree contributions, such that all non-trivial strong-interaction
phases are accounted for by the phase difference $\Delta\phi$.

The value of the parameter $\varepsilon_{3/2}$ can be estimated using
various experimental and theoretical information. The penguin
contribution $|P_{ct}|$ can be  extracted from the CP-averaged 
branching ratio for the decays $B^\pm\to\pi^\pm K^0$, which are
expected to receive only a very small contamination from other
contributions such as up-quark penguins or annihilation diagrams.
Using the experimental value for this branching ratio
quoted earlier yields $|P_{ct}|=(3.74\pm 0.72)\times 10^{-3}$. (We
give amplitudes  in ``branching ratio units'', so that
$\mbox{Br}=|{\mathcal A}|^2$.) The tree contribution $|T+C|$ can be
estimated experimentally invoking SU(3) symmetry, in which case it can
be extracted from the branching ratio for the decay
$B^\pm\to\pi^\pm\pi^0$ using the relation \cite{GRL}--\cite{Zepp}
\begin{equation}
   T+C = -\sqrt 2\,\frac{V_{us}}{V_{ud}}\,\frac{f_K}{f_\pi}\,
   {\mathcal A}(B^+\to\pi^+\pi^0) \,,
\label{SU3rel}
\end{equation}
where the factor $f_K/f_\pi=1.22\pm 0.01$ accounts for the leading
(i.e., factorizable) SU(3)-breaking corrections. The CLEO
Collaboration  has recently reported a signal for this decay mode with
a significance  of 2.3 standard deviations. They quote the upper limit
$\mbox{Br}(B^\pm\to\pi^\pm\pi^0)<1.6\times 10^{-5}$ at 90\% confidence
level \cite{CLEOnew}, which implies $|T+C|<1.56\times 10^{-3}$.
Taking this signal seriously, we may derive from the reported event
rate and efficiency the branching ratio
$\mbox{Br}(B^\pm\to\pi^\pm\pi^0)=(0.59_{-0.27}^{+0.32})\times
10^{-5}$, which yields the value  $|T+C|=(0.94_{-0.21}^{+0.25})\times
10^{-3}$.  The tree contribution may also be obtained theoretically
employing the factorization hypothesis. Using the values of the
hadronic form factors determined in Ref.~\cite{Stech} combined with
the new world average $|V_{ub}|=(3.56\pm 0.56)\times 10^{-3}$
\cite{Rosnet}, we get $|T+C|=(0.88\pm 0.23)\times 10^{-3}$, in good
agreement with the  estimate obtained using experimental data.  We
take the average of the two numbers and combine it with the
experimentally determined value for the penguin amplitude to obtain
$\varepsilon_{3/2}=0.24\pm 0.06$. It is important that, once the 
branching ratio for the decays $B^\pm\to\pi^\pm\pi^0$ and 
$B^\pm\to\pi^\pm K^0$ are measured more precisely, the parameter 
$\varepsilon_{3/2}$ can be determined with smaller errors and without 
almost any recourse to theory. Note that with the above value of 
$\varepsilon_{3/2}$ the terms of $O(\varepsilon^2)$ omitted in  
(\ref{Delstar}) are expected to be reasonably small compared with the 
sizable experimental uncertainties in (\ref{Rstexp}). Nevertheless, 
we shall comment on the relevance of such higher-order terms below.

Inserting the value of $\varepsilon_{3/2}$ into relation 
(\ref{Delstar}) and neglecting higher-order terms, we obtain
\begin{eqnarray}
   \cos\Delta\phi\,(\delta_{\rm EW} - \cos\gamma)
   &=& \frac{\Delta_*}{\varepsilon_{3/2}} = 1.33\pm 0.78 \nonumber\\
   \Rightarrow \quad |\delta_{\rm EW} - \cos\gamma|
   &\ge& 1.33\pm 0.78 \,.
\label{newbound}
\end{eqnarray}
The last step missing to turn this result into a useful constraint on
$\gamma$ is an estimate of the contribution from electroweak  penguins
to the isospin amplitude $A_{3/2}$. This can be done  without
encountering hadronic uncertainties. The crucial observation  is that
the electroweak penguin operators $Q_9$ and $Q_{10}$, whose Wilson
coefficients are enhanced by the large mass of the top quark, are
Fierz-equivalent to the current--current operators $Q_1$ and $Q_2$.
As a result, the $\Delta I=1$ part of the effective weak Hamiltonian
for $B\to\pi K$ decays can be written as \cite{Ne97,Robert}
\begin{equation}
   {\mathcal H}_{\Delta I=1} = \frac{G_F}{\sqrt 2} \left\{ \left(
   \lambda_u C_1 - \frac 32\lambda_t C_9 \right) \bar Q_1 + \left(
   \lambda_u C_2 - \frac 32\lambda_t C_{10} \right)  \bar Q_2 + \dots
   \right\} + \mbox{h.c.} \,,
\label{newH}
\end{equation}
where $\bar Q_i=\frac 12(Q_i^u-Q_i^d)$.  The dots represent the
contributions from the electroweak penguin operators $Q_7$ and $Q_8$,
which have a different Dirac structure. The Wilson coefficients of
these operators are so small that their contributions can be safely
neglected.  It is important in this context that for heavy mesons the
matrix elements of four-quark operators with Dirac structure
$(V-A)\otimes (V+A)$ are not enhanced with respect to those of
operators  with the usual $(V-A)\otimes(V-A)$ structure, as can be
checked explicitly in the  factorization approximation. To an
excellent approximation, the net effect of electroweak penguin
contributions to the  $\Delta I=1$ isospin amplitudes in $B\to\pi K$
decays thus consists of the replacements of the Wilson coefficients
$C_1$ and $C_2$ of the current--current operators with the
combinations shown in (\ref{newH}). In the SU(3) limit, $U$-spin
invariance implies that the isospin amplitude $A_{3/2}$ is given by
the matrix element of the combination $(\bar Q_1+\bar Q_2)$ only, the 
coefficient of which is $\frac 12[\lambda_u
(C_1+C_2)-\frac 32\lambda_t(C_9+C_{10})]$.  The difference, $(\bar
Q_1-\bar Q_2)$, does not contribute to this  amplitude. To see this,
consider the relation (\ref{SU3rel}) between the current--current part
of the amplitude $A_{3/2}$ and the amplitude for the decay
$B^+\to\pi^+\pi^0$. Bose symmetry requires the two pions to be in an
$I=2$ state, so that only the $\Delta I=\frac 32$ part of the
effective weak Hamiltonian for $B\to\pi\pi$ decays contributes to the
decay. Since the combination $(\bar Q_1'-\bar Q_2')$, where the prime
on the operators indicates the replacement $s\to d$  appropriate for
$\Delta S=0$ transitions, has $\Delta I=\frac 12$, the decay
$B^+\to\pi^+\pi^0$ receives a contribution from the combination $(\bar
Q_1'+\bar Q_2')$ only. SU(3) symmetry then implies that it is only the
combination  $(\bar Q_1+\bar Q_2)$ which contributes to the isospin
amplitude $A_{3/2}$ in $B\to\pi K$ decays. It follows that the
quantity  $\delta_{\rm EW}$ in (\ref{ratio}) is given by
\begin{equation}
   \delta_{\rm EW} = - \frac{3}{2\lambda^2 R_b}
   \frac{C_9+C_{10}}{C_1+C_2} = 0.66\pm 0.11 \,,
\label{dEW}
\end{equation}
where we have used $\lambda_u/\lambda_t\approx-\lambda^2 R_b\,
e^{i\gamma}$ with $\lambda\approx 0.22$ and $R_b=\lambda^{-1}
|V_{ub}/V_{cb}|\approx 0.41\pm 0.07$. The ratio of the Wilson
coefficients is to a very good approximation independent of the choice
of the renormalization scale.  We take $(C_9+C_{10})/(C_1+C_2)=
-1.14\alpha$, which is obtained using the leading-order
coefficients\footnote{Using next-to-leading order coefficients would
change this value by less than 2\%; however, at this order  the
coefficient functions may depend on the choice of the operator basis,
so one has to be careful when using Fierz indentities.}  at the scale
$\mu=m_b$ \cite{Heff}.  For the electromagnetic coupling at the weak
scale we take  $\alpha=1/129$.  Note that the main uncertainty in the
value of $\delta_{\rm EW}$ results from the uncertainty in $V_{ub}$,
which is likely to be reduced in the near future.

SU(3)-breaking  corrections to the relation (\ref{dEW}) are controlled
by the ratio of the following operator matrix elements:
\begin{equation}
   \frac{\langle \pi K(I=\textstyle{\frac 32})|\,\bar Q_1-\bar Q_2\,
|B^+\rangle}{\langle \pi K(I=\textstyle{\frac 32})|\, \bar Q_1+\bar
Q_2\,|B^+\rangle}  \equiv - \delta_{\rm SU(3)}\,e^{i\Delta\varphi} \,.
\end{equation}
The magnitude of this ratio can be estimated using the generalized
factorization hypothesis  \cite{Stech} to calculate the matrix
elements of the current--current  operators. This leads to
\begin{equation}
   \delta_{\rm SU(3)} = \frac{1-\zeta}{1+\zeta}\,\frac{a_K -
   a_\pi}{a_K + a_\pi} \approx 1\mbox{--}3\% \,,\qquad \Delta\varphi =
   0 \,,
\end{equation}
where $a_K=f_K (m_B^2-m_\pi^2) F_0^{B\to\pi}(m_K^2)$ and  $a_\pi=f_\pi
(m_B^2-m_K^2) F_0^{B\to K}(m_\pi^2)$ are combinations of hadronic
matrix elements, and $\zeta\approx 0.45$ is a parameter  controlling
non-factorizable corrections. Despite the fact that the factorization
approximation may be crude, this estimate suggests that SU(3)-breaking
corrections to the result (\ref{dEW}) are very small. To linear order
in SU(3) breaking, the results (\ref{Delstar})  and (\ref{newbound})
can be corrected by replacing $\delta_{\rm EW}$ with the effective
value
\begin{equation}
   \delta_{\rm EW}^{\rm eff} = \left( 1 - k\,\delta_{\rm SU(3)}\,
   \frac{\cos(\Delta\phi+\Delta\varphi)}{\cos\Delta\phi} \right)
   \delta_{\rm EW} \,,
\end{equation}
where
\begin{equation}
   k = \frac{C_2-C_1}{C_2+C_1} + \frac{C_9-C_{10}}{C_9+C_{10}} \approx
   3.43
\end{equation}
is a combination of Wilson coefficients. Since $(\bar Q_1-\bar Q_2)$
and $(\bar Q_1+\bar Q_2)$ are local operators, it is unlikely that the
strong-interaction phase difference $\Delta\varphi$ could acquire any
significant value. We thus expect that the main effect of SU(3)
breaking is to reduce the  value of $\delta_{\rm EW}$ in (\ref{dEW})
by an amount of order 5\%. To account for this effect, we  shall from
now on use the value  $\delta_{\rm EW}=0.63\pm 0.15$ with an increased
error, which is large enough to cover possible small contributions
from a non-zero phase difference $\Delta\varphi$ or deviations from the
factorization approximation. It is interesting to note that our
general results for the structure of the electroweak penguin
contributions to the isospin amplitude $A_{3/2}$, including the
pattern of SU(3)-breaking effects, are in full accord with  a model
estimate by Deshpande and He \cite{DeHe}. However, these authors did
not realize that such contributions can be
calculated in an essentially model-independent way. As a result,
electroweak penguin effects can be taken into account in the method
proposed in Ref.~\cite{GRL} for learning $\gamma$ by comparing the
rates for the processes $B^+\to\pi^+ K^0$, $\pi^0 K^+$, and
$\pi^+\pi^0$ with their charge conjugates. This will be discussed in
more detail elsewhere \cite{newpaper}.

\begin{table}
\centerline{\parbox{14cm}{\caption{\label{tab:1} 
Constraints on $\gamma$ for different values of the ratio 
$\Delta_*/\varepsilon_{3/2}$}}}
\begin{center}
\begin{tabular}{||c|c|c|c||}
\hline\hline  
$\Delta_*/\varepsilon_{3/2}$ & deviation from central value &
 bound on $\cos\gamma$ & excluded region \\  
\hline\hline   
2.11 & $+1\sigma$ & no solution & all \\   
1.33 & central value & $<-0.55$ & $|\gamma|<123^\circ$ \\  
0.55 & $-1\sigma$ & $<0.23$ & $|\gamma|<77^\circ$ \\   
0.16 & $-1.5\sigma$ & $>0.64\,$ or $\,<0.62$ &
 $50^\circ<|\gamma|<52^\circ$ \\   
\hline\hline
\end{tabular}
\end{center}
\end{table}

The role of electroweak penguins in the bound (\ref{newbound}) is an
important one, and it is crucial that the parameter $\delta_{\rm EW}$
can be calculated in a reliable way in terms of perturbative Wilson
coefficients and measured quantities. The result is that the
constraint (\ref{newbound}) excludes values of $|\gamma|$ around
$\mbox{arccos}(0.63)\approx 51^\circ$.  More precisely, the excluded
region depends on the value of the experimental number for the ratio
$\Delta_*/\varepsilon_{3/2}$ on the right-hand side of the bound.  The
allowed regions for $\cos\gamma$ must satisfy  $\cos\gamma>\delta_{\rm
EW}+\Delta_*/\varepsilon_{3/2}$ or $\cos\gamma<\delta_{\rm
EW}-\Delta_*/\varepsilon_{3/2}$. In evaluating these results we lower 
or increase the value of $\delta_{\rm EW}$ by one standard deviation 
so as to be conservative, i.e., we use $\delta_{\rm EW}=0.48$ in the
first relation and $\delta_{\rm EW}=0.78$ in the second one. The first
solution only exists if $\Delta_*/\varepsilon_{3/2}<1-\delta_{\rm EW}$.  
If $\Delta_*/\varepsilon_{3/2}>1+\delta_{\rm EW}$ there is no solution 
for $\gamma$ at all. The results obtained for some representative 
values of $\Delta_*/\varepsilon_{3/2}$ are shown in Table~\ref{tab:1}.  
It is remarkable that the constraints on $\gamma$ derived from the bound
(\ref{newbound}) are to a large extent complementary to the preferred
region for this angle obtained from a global analysis of the unitarity
triangle, using information from semileptonic $B$ decays, $B$--$\bar
B$ mixing, and CP violation in the kaon system.  A typical range of
values allowed by such an analysis is $47^\circ<\gamma<105^\circ$
\cite{Jonnew}, where the upper bound is determined by the improved
lower limit on the quantity $\Delta  m_s$ controlling 
$B_s$--$\bar B_s$ mixing \cite{CLEOnew}.  For the central value of
$\Delta_*/\varepsilon_{3/2}$ our bound is inconsistent with this region,
while for the $1\sigma$ lower value $\Delta_*/\varepsilon_{3/2}=0.55$ it
would exclude values of $|\gamma|$ below $77^\circ$, thus leaving the
rather narrow range $77^\circ<\gamma<105^\circ$. Clearly, a more
precise measurement of the parameter $\Delta_*/\varepsilon_{3/2}$ could
potentially yield a very non-trivial constraint on $\gamma$ and
provide a stringent test of the CKM paradigm.

The weakest point of our analysis appears to be the
neglect of higher-order terms in $\varepsilon$ in the
expression  for the quantity $\Delta_*$ in (\ref{Delstar}), given that
$\varepsilon_{3/2}\approx 0.24$ is not such a small parameter.  
We recall that $\varepsilon$ is a generic measure of the strength of the
Cabibbo-suppressed $\bar b\to\bar u u\bar s$ transitions, which carry
the weak phase $e^{i\gamma}$, relative to the leading  penguin
transitions.  Whereas a full account of all higher-order terms of this
kind would  introduce unknown strong-interaction parameters, we now
show that the potentially most important  terms can be analysed
without much additional complication.  The reason is that
contributions proportional to the weak phase $e^{i\gamma}$ in the
decay amplitude for $B^+\to\pi^+ K^0$ are likely to be much smaller
than those in the amplitude for $B^+\to\pi^0 K^+$, because they could
only be induced via final-state rescattering through soft annihilation
or up-quark penguin topologies \cite{Ne97}. Model estimates indicate
that those effects typically lead to contributions of order a few
percent \cite{GRKpi}--\cite{At97}. In other words, if we write 
$B_{1/2}+A_{1/2}+A_{3/2}=|P_{ct}| e^{i\phi_P} [e^{i\pi} + e^{i\gamma} 
e^{i\eta} \varepsilon_a]$, we expect that $\varepsilon_a\ll
\varepsilon_{3/2}$. This observation justifies treating these 
rescattering contributions from a numerical point of view as quadratic 
rather than linear in $\varepsilon_{3/2}$. The generalization of the 
relation (\ref{Delstar}) to the next order in $\varepsilon_{3/2}$ then 
becomes
\begin{eqnarray}
   \Delta_* &=& \varepsilon_{3/2} \cos\Delta\phi\,
    (\delta_{\rm EW} - \cos\gamma) \nonumber\\
   &&\mbox{}- \frac{\varepsilon_{3/2}^2}{2}
    \left[ (3\cos^2\!\Delta\phi - 1) (\delta_{\rm EW} - \cos\gamma)^2
    - \sin^2\!\gamma \right] 
    + O(\varepsilon_{3/2}^3,\varepsilon_a\varepsilon_{3/2}) \,.
\end{eqnarray}
The question is whether the terms of order $\varepsilon_{3/2}^2$ could
increase the value of $\Delta_*$, thereby weakening the bound on
$\gamma$ derived in this letter. Since $\varepsilon_{3/2}\ll 1$, the
quantity $\Delta_*$ takes its maximum value for $|\cos\Delta\phi|=1$,
so that
\begin{equation}
   \Delta_* \le \varepsilon_{3/2}\,|\delta_{\rm EW} - \cos\gamma|
   - \left(\Delta_*^2 - \frac{\varepsilon_{3/2}^2}{2} \sin^2\!\gamma
   \right) + O(\varepsilon_{3/2}^3,\varepsilon_a\varepsilon_{3/2}) \,.
\end{equation}
If experimentally it is found that $\Delta_*>\varepsilon_{3/2}/\sqrt 2
\approx 0.17$, the second term is negative and strengthens the bound
on $\gamma$. Provided future measurements confirm the present value of
$\Delta_*$ in (\ref{Rstexp}) within one standard deviation, the bound
(\ref{newbound}) derived in linear approximation is thus a
conservative result. On the other hand, if $\Delta_*$ would turn out
to be much smaller than the current central value, the bound may be
weakened by higher-order terms, but those would generally be smaller
than $\varepsilon_{3/2}^2/2\approx 3\%$, which is an almost negligible
amount.

In summary, we have shown that a non-trivial bound on the angle
$\gamma$ of the unitarity triangle can be derived from a measurement
of the branching ratios for the  decays $B^\pm\to\pi K$, averaged over
CP-conjugate modes.  Electroweak penguin contributions, which play an
important role in these decays, can be controlled in a
model-independent way using Fierz identities and SU(3) symmetry 
relations, with strong indications that SU(3)-breaking corrections are 
very small.

\vspace{0.3cm} 
{\it Acknowledgments:\/} 
Part of this work was done during the Workshop on {\sl Perturbative
and Non-Perturbative Aspects of the Standard Model\/} at St.\ John's
College, Santa Fe, July--August 1998.  We would like to thank the
organizer Rajan Gupta, as well as the participants of the workshop,
for providing a stimulating atmosphere and for many useful
discussions.  We are grateful to Michael Gronau, Alex Kagan, Sheldon 
Stone and Lincoln Wolfenstein for helpful comments on the manuscript. 
One of us (J.L.R.) was supported in part by the United States 
Department of Energy through contract No.\ DE FG02 90ER40560.

\end{document}